%% file: main.tex
\documentclass[a4paper,UKenglish]{lipicsmaster-v2021}

\hideLIPIcs
\nolinenumbers
\usepackage[utf8]{inputenc}
\usepackage{microtype}
\usepackage{mathtools}
\usepackage{booktabs}
\usepackage{xurl}
\usepackage{tikz}
\usetikzlibrary{positioning,arrows.meta,calc}

\newcommand{\OPT}{\operatorname{OPT}}
\newcommand{\LP}{\operatorname{LP}}
\newcommand{\cost}{\operatorname{cost}}
\newcommand{\SC}{\operatorname{SC}}

\title{Individually Fair Facility Location\footnote{We thank Adobe Research Labs for gift funding and NSF grant 2216970.}}
\titlerunning{Individually Fair Facility Location}

\author{Rajni Dabas}
       {Northwestern University}
       {rajni.dabas1@northwestern.edu}
       {}
       {}
\author{Samir Khuller}
       {Northwestern University}
       {samir.khuller@northwestern.edu}
       {}
       {}

\author{Milind Maiti}
       {Northwestern University}
       {milindmaiti2027@u.northwestern.edu}
       {}
       {}

\authorrunning{R. Dabas, S. Khuller, and M. Maiti}

\Copyright{Rajni Dabas, Samir Khuller, and Milind Maiti}


\EventEditors{}
\EventLongTitle{}
\EventShortTitle{FORC}
\EventAcronym{FORC}
\EventYear{}
\EventDate{}
\EventLocation{}
\EventLogo{}
\SeriesVolume{}
\ArticleNo{}

\AtBeginDocument{\hypersetup{
  hidelinks,
  pdftitle={Individually Fair Facility Location},
  pdfauthor={},
  pdfsubject={FORC submission}
}}

\begin{document}

\maketitle

\input{abstract}
\input{intro}
\input{hardness}

\input{1.725lp}

\input{exp-rewritten}
\input{Conclusion}

\bibliography{references}

\appendix
\input{derandomization}
\input{badExampleforDualFitting}

\end{document}

%% file: abstract.tex
\begin{abstract}
We study individually fair facility location, where each client must connect to an open facility within a prescribed radius and the objective is to minimize total opening and connection cost. We show that preserving the radii changes the approximability of metric facility location fundamentally. For every fixed radius violation below three, no deterministic polynomial-time algorithm achieves a sublogarithmic cost factor unless $\mathrm{P}=\mathrm{NP}$. The hardness holds with uniform opening costs and uniform radii. It contradicts the strict-fairness constant-factor guarantee recently claimed by Liu, Xu, and Zhang [{\em Theory of Computing Systems}, Vol. 70 (2026)]. 

We then establish a positive result for arbitrary client radii. We give a randomized polynomial-time algorithm that has cost at most $1.725$ times the strictly fair optimum and serves every client within three times its fair radius (in the Appendix we show how this can be de-randomized). We use the approach of Byrka, Srinivasan and Swamy [{\em IPCO 2010}]. The facility-location radius factor {\em three} is therefore optimal among deterministic polynomial-time algorithms with constant cost factor, unless $\mathrm{P}=\mathrm{NP}$. Experiments on real-world data indicate that the worst case radius bound is rarely realized in practice: the algorithm serves over 99.8\% of clients within their stated radius, and never beyond 1.5 times it. Our code is available at \url{https://github.com/milindmaiti/IFFL_Implementation}.


\end{abstract}

%% file: intro.tex
\section{Introduction}
In the classical facility location (FL) problem, given a set of clients
and a set of potential facilities in a metric space. Each facility has
an opening cost, and the cost of serving a client is the distance to the
facility serving it. The objective is to choose which facilities to open
and assign every client to an open facility so as to minimize the total
opening and service cost.  This problem has been widely studied in the area of approximation algorithms starting from  Hochbaum \cite{Hochbaum82} and Shmoys et al \cite{ShmoysTardosAardal1997}. 
The best approximation ratio of \(1.488\) for FL was given by Li \cite{Li2013}, while approximating within any factor below \(1.463\) is NP-hard \cite{GuhaKhuller1999}.

Unlike in the K-center problem where we guarantee fairness in a way (by minimizing the maximum distance any client has to travel), in problems such as K-medians or FL there is no constraint on the distance a client has to travel. Some clients might travel a short distance, and others might travel very far to get to a facility. Individual fairness constraints add a local requirement: each client has a service radius and must find an open facility within that radius (referred to as its fair radius). When radii are determined by local density, clients in dense regions may require closer service than clients in sparse regions. This interpretation follows the neighborhood-based view of individual fairness introduced by Jung, Kannan, and Lutz~\cite{JungKannanLutz2020}.

Individual fairness has been studied extensively for clustering objectives such as $k$-median\footnote{Instead of facility opening costs, a hard bound $k$ is given on number of facilities that can be opened in the final solution.}, where approximation algorithms trade
connection cost against a multiplicative relaxation of the radii \cite{AlamdariShmoys2018}. The corresponding tradeoff for facility location requires separate
analysis. Deleting every client-facility edge that violates its client's radius gives an exact formulation of the feasible solutions, but the usual metric approximation guarantees do not follow from this deletion. Facility-location algorithms may reassign a client along a path through other clients and facilities. Although that path is short in the underlying metric, its endpoint need not satisfy the original client's
radius constraint. As we will show in this paper, the admissible edges can encode an arbitrary Set Cover instance, even when every radius is the same and the underlying distances belong to $\{1,2,3\}$.

This issue is central to the recent work of Liu, Xu, and Zhang~\cite{LiuXuZhang2026}, which claimed a polynomial-time $2$-approximation for individually fair facility location while
preserving every radius exactly. We show that such a guarantee is impossible unless $\mathrm{P}=\mathrm{NP}$. More generally, even a
sublogarithmic cost approximation is impossible for any fixed radius violation below three. At radius factor three, however, we obtain a deterministic $1.725$-factor in cost for arbitrary client radii. 

\subsection{Problem definition and our results}
Let $F$ be a finite set of candidate facilities and $C$ a finite set of
clients in a common metric space with distance function $d$. Let $n=|C|$ and $m=|F|$. Each facility $i\in F$ has an opening cost $f_i\ge0$, and each client $j\in C$ has a fair radius $R_j\ge0$. Facilities and clients may be distinct sets. For a nonempty set $S\subseteq F$, define
\[
 d(j,S)=\min_{i\in S}d(i,j),
 \qquad
 \operatorname{cost}(S)
 =\sum_{i\in S}f_i+\sum_{j\in C}d(j,S).
\]

\begin{definition}[Individually fair facility location]
A solution $S\subseteq F$ is fair if $d(j,S)\le R_j$ for every
client $j\in C$. Let $\mathrm{OPT}_{\mathrm{FL}}$ denote the minimum
cost of a strictly fair solution.
\end{definition}

We consider feasible instances: every client has at least one candidate facility within its radius. We first prove that a radius relaxation of at least three is necessary for deterministic polynomial-time constant-cost approximation.


\begin{theorem}\label{thm:hardness-intro}
Fix $1\leq\beta<3$. Unless $\mathrm{P}=\mathrm{NP}$, no deterministic polynomial-time $(\rho(n),\beta)$-approximation exists for $\rho(n)=o(\log n)$. This holds on feasible instances with equal positive facility costs, all nonzero distances in $\{1,2,3\}$, and $R_j=1$ for every client. 
\end{theorem}

The proof encodes Set Cover in the admissible client-facility edges. In the constructed metric, incident facilities are at distance one and non-incident facilities are at distance three. Any radius violation strictly below three must therefore retain the set-cover structure. Theorem~\ref{thm:hardness-intro} rules out the claimed $(2,1)$ guarantee of Liu, Xu, and Zhang\cite{LiuXuZhang2026}. We also give an explicit counterexample to their algorithm and proposed dual scaling. 

We next present a bi-criteria approximation for the problem with arbitrary radii using LP rounding technique as stated in Theorem \ref{thm:general-intro}. A bi-criteria approximation is defined as follows.

\begin{definition}[Bicriteria approximation]
A $(\rho,\beta)$-approximation returns $S\subseteq F$ satisfying
\[
 \cost(S)\leq\rho\OPT_{\rm FL},\qquad
 d(j,S)\leq\beta R_j\quad(j\in C).
\]
The comparison is always with the optimum satisfying the \emph{original} radii, rather than the optimum for the relaxed radii.
\end{definition}
Our algorithm attains the endpoint radius factor while preserving a
cost factor below two.
\begin{theorem}\label{thm:general-intro}
There is a deterministic polynomial-time $(1.725,3)$-approximation for facility location with arbitrary client radii.
\end{theorem}

Byrka, Srinivasan, and Swamy~\cite{ByrkaSrinivasanSwamy2010} give a $1.725$-approximation for fault-tolerant facility location, where a client may require connections to several distinct facilities. Their rounding processes clients in nondecreasing order of the
maximum distance in their close fractional support and guarantees
nearby backup facilities. Our upper bound applies this locality
principle to the radius restricted LP with one required connection per client. The original metric is retained, and the restriction on fractional assignments ensures that each client's close support lies
within its prescribed radius. Section \ref{sec:rounding} gives a direct proof of this specialization using disjoint rounding groups, an elementary probability comparison, and an explicit derandomization. The upper bound establishes attainability of radius factor three; together with our lower bound, it determines the radius threshold for constant-cost approximation.

That threshold is a worst-case guarantee, and Section~\ref{sec:experiments}
examines how much of it is needed in practice. We compare the rounding of
Theorem~4 with two reference algorithms: a greedy algorithm restricted to admissible pairs, which is strictly fair
by construction, and a Jain--Vazirani style primal-dual algorithm. Our rounding
exceeds the restricted LP optimum by $1.8\%$ on average and serves over
$99.8\%$ of clients within their prescribed radius rather than the guaranteed
$3R_j$; greedy pays $1.4\%$ over the same bound. The primal-dual algorithm,
whose $(3,3)$ guarantee holds only for uniform radii, violates the radius for
about a fifth of clients once radii vary, reaching $216\,R_j$ ---the failure an
arbitrary-radius guarantee rules out.



\subsection{Related work and comparison}
We report bicriteria guarantees in the order $(\text{cost},\text{radius})$, with cost compared against the optimum satisfying the original radii.

Individual fairness based on neighborhood radii was introduced by Jung, Kannan, and Lutz~\cite{JungKannanLutz2020}. The simultaneous minimization of clustering cost under these constraints has subsequently been studied
extensively, including for the $k$-median objective. For individually fair $k$-median in general metrics, Mahabadi and Vakilian~\cite{MahabadiVakilian2020} give a $(84,7)$-approximation using local search; Chakrabarty and Negahbani~\cite{ChakrabartyNegahbani2021} give an $(8,8)$-approximation using LP rounding; and Vakilian and Yal\c{c}{\i}ner~\cite{VakilianYalciner2022} obtain a $(7.081+\varepsilon,3)$-approximation through matroid-constrained clustering. In metrics of fixed doubling dimension, Wu, Feng, and Wang~\cite{WuFengWang2026} give a randomized $(1+\varepsilon,2+\varepsilon)$-approximation. This last guarantee uses the additional geometric assumption and should be distinguished from
general-metric results. 

A closely related line of work minimizes $k$-median cost subject to a common radius bound. Alamdari and Shmoys~\cite{AlamdariShmoys2018} give an $(8,4)$-approximation. For this common-radius problem, Chi, Guo, and Jia~\cite{ChiGuoJia2025} give a $(3+\varepsilon,7)$-approximation by local search, and Arutyunova et al.~\cite{ArutyunovaEtAl2024} obtain a
$(6+\varepsilon,9)$-approximation. 

For facility location problem and common client radii $R_j=R$, a $(3,3)$ guarantee follows from
the adaptation of Jain and Vazirani's primal-dual method~\cite{JainVazirani2001} given by Arutyunova
et al.~\cite{ArutyunovaEtAl2024}. Their cost analysis gives a factor of three, while an indirectly connected client is reassigned along three edges, each of length at most $R$, giving a service distance of at most $3R$. This is a known guarantee for the common-radius setting. Bajpai and Chekuri~\cite{BajpaiChekuri2023} study priority matroid median, which combines arbitrary client radii and facility opening costs with a matroid constraint on the open facilities. They obtain $(12,21)$- and $(8,36)$-approximations in our $(\text{cost},\text{radius})$ notation.

To the best of our knowledge, the primary  work on Fair facility location is that of Liu, Xu, and Zhang~\cite{LiuXuZhang2026}.
They claim a $(2,1)$-approximation for facility location and a $(4,2)$ guarantee for individually fair $k$-median via their facility-location subroutine. Theorem~\ref{thm:hardness-intro} rules out their claimed facility-location guarantee unless $\mathrm{P}=\mathrm{NP}$; our explicit
counterexample also shows a failure of the proposed algorithm and dual scaling. 

\textbf{Other notions of fairness in facility location.}
Fair clustering can constrain the proportion of each group within a
cluster~\cite{ChierichettiEtAl2017,BeraEtAl2019,BerceaEtAl2019,CaselEtAl2023}.
Other models incorporate group-average service costs into the
objective~\cite{AbbasiEtAl2021,GhadiriEtAl2021,GuptaMoondraSingh2023}.
Related work examines how differential privacy can affect groups or individuals unequally in resource allocation and facility placement~\cite{PujolEtAl2020,FishEtAl2026}.
When some clients may be left unserved, fair-outlier models bound the
number of exclusions in each
group~\cite{InamdarVaradarajan2018,AlmanzaEtAl2022,ChekuriEtAl2022,BajpaiChekuriKulkarni2025,DabasKhullerRivkin2026}.

\subsection{Organization}

Section~2 proves the hardness result. Section~3 presents the LP-rounding
algorithm and its cost and radius analysis. The appendices give the derandomization details and the explicit counterexample to the previously claimed strict-fairness approximation.

%% file: hardness.tex
\section{Hardness of Individually Fair Facility Location}\label{sec:hardness}
We prove the hardness using a reduction from the unweighted set cover problem. Let $U=\{u_1,\ldots,u_q\}$ and $\mathcal S=\{S_1,\ldots,S_p\}$ be an unweighted Set Cover instance with $q\geq1$ and $\bigcup_{\ell=1}^p S_\ell=U$. Let $k^*=\OPT_{\SC}$ be its minimum cover size. Create two\footnote{Each client’s nearest other client is its partner at distance \(1\), so the definition in Liu et al. \cite{LiuXuZhang2026} gives every client fair radius \(1\) with \(\tau=2\). When fair radii are given as input, one client per element suffices.} clients $j_{a0},j_{a1}$  for each element $u_a$ and one facility $i_\ell$ for each set $S_\ell$. Hence $n=|C|=2q$. Every facility has opening cost $f_i=M=q^2$. Set $R_j=R=1$ for all $j$. Define $d(v,v)=0$
and define the remaining distances symmetrically as follows. Note that all the distances between distinct points lie in $\{1,2,3\}$ and form a metric.
\begin{align}
d(j_{a0},j_{a1})&=1,\quad
d(j_{ab},j_{cb'})=2\ (a\ne c),\quad
d(i_\ell,i_h)=2\ (\ell\ne h),\nonumber\\
d(j_{ab},i_\ell)&=
\begin{cases}
1,&u_a\in S_\ell,\\
3,&u_a\notin S_\ell.
\end{cases}
\label{eq:hard-metric}
\end{align}


\begin{lemma}\label{lem:cover-cost}
The constructed instance is strictly feasible, $\OPT_{\rm FL}=Mk^*+2q.$
Any solution with service distances strictly below three opens facilities corresponding to a set cover.
\end{lemma}
\begin{proof}
Every element belongs to some set. Its two clients therefore have an incident facility at distance one, which proves feasibility. A set cover of size $k$ gives a strictly fair solution of cost $Mk+2q$. Conversely, every client--facility distance is either one or three. A strictly fair solution, or any solution with service distances below three, can use only incidence edges. Its open sets must cover every element, and each connection costs exactly one. The minimum number of open facilities is therefore $k^*$.
\end{proof}

\begin{proof}[Proof of Theorem~\ref{thm:hardness-intro}]
Suppose an algorithm has cost factor $\rho(n)$ and radius factor $\beta<3$. Apply it to the constructed instance and let $k_{\rm alg}$ be the number of opened facilities. By Lemma~\ref{lem:cover-cost}, these facilities give a set cover. The cost guarantee implies $
 Mk_{\rm alg}\leq\rho(2q)(Mk^*+2q).$ Since $M=q^2$ and $k^*\geq1$,
\begin{equation}\label{eq:cover-transfer}
 \frac{k_{\rm alg}}{k^*}
 \leq\rho(2q)\left(1+\frac{2}{qk^*}\right)
 \leq\left(1+\frac2q\right)\rho(2q).
\end{equation}
The metric has $2q+p$ points and polynomial encoding length. If $\rho(n)=o(\log n)$, equation~\eqref{eq:cover-transfer} gives a polynomial-time $o(\log q)$-approximation for Set Cover. This contradicts its logarithmic inapproximability unless $\mathrm{P}=\mathrm{NP}$~\cite{DinurSteurer2014,Moshkovitz2015}.

For the sharper statement, fix $\varepsilon\in(0,1)$ and take $\rho(n)=(1-\varepsilon)\ln n$. For all sufficiently large $q$,
$
 \left(1+\frac2q\right)(1-\varepsilon)\ln(2q)
 <(1-\varepsilon/2)\ln q,
$
contradicting the corresponding Set Cover hardness. Any finitely many smaller universe sizes can be solved exactly in polynomial time by dynamic programming over their subsets. This proves the claimed asymptotic approximation impossibility.
\end{proof}

At radius factor exactly three, a client may use a nonincident facility, so the cover extraction no longer holds. Section~\ref{sec:rounding} shows that this endpoint is achievable with a constant cost factor. The complexity assumption in Theorem~\ref{thm:hardness-intro} concerns deterministic algorithms; we make no claim that $\mathrm{P}\ne\mathrm{NP}$ alone rules out randomized expected-cost approximations.


%% file: 1.725lp.tex
\section{Facility location with arbitrary radii}\label{sec:rounding}

We now give an algorithm for facility location with arbitrary client radii. Its cost is compared with the optimum satisfying the original radii,
whereas the algorithm is allowed to increase every radius by a factor of three. We assume feasibility which means that each client has at least one facility
within its prescribed radius. The algorithm uses the scaling, facility-splitting, and cluster-rounding approach developed for facility location
\cite{ChudakShmoys2003,ByrkaAardal2010,ByrkaSrinivasanSwamy2010}. We give a self-contained cost analysis and derandomize the resulting algorithm. Details of derandomization appear in Appendix \ref{sec:deram}. Throughout the section, fix $\gamma=1.725$.

\subsection{A restricted LP and filtered facility sets}
Let $E_R=\{(i,j)\in F\times C:d(i,j)\le R_j\}$. We solve the following restricted facility-location relaxation
\begin{equation}\label{eq:restricted-lp}
\begin{aligned}
 \LP=\min\quad &\sum_{i\in F} f_i y_i
                  +\sum_{(i,j)\in E_R}d(i,j)x_{ij}\\
 \text{subject to}\quad
 &\sum_{i:(i,j)\in E_R}x_{ij}=1 &&(j\in C),\\
 &0\le x_{ij}\le y_i &&((i,j)\in E_R),\\
 &0\le y_i\le1 &&(i\in F).
\end{aligned}
\end{equation}
For an optimal solution $(x^*,y^*)$, let
\[
 F^*=\sum_i f_i y_i^*,\qquad
 C_j^*=\sum_{i:(i,j)\in E_R}d(i,j)x_{ij}^*,\qquad
 C^*=\sum_j C_j^*.
\]
Thus $F^*+C^*=\LP\le\OPT_{\rm FL}$.

We complete the fractional solution by splitting facilities. We briefly recall the standard facility-splitting procedure for obtaining a complete fractional solution \cite{ChudakShmoys2003}.  Whenever \(0<x_{ij}^*<y_i^*\), let \(\delta=\min_{k:x_{ik}^*>0}x_{ik}^*\). Transfer \(\delta\) of \(i\)'s opening and of each of its positive assignments to a copy at the same location with the same opening cost. Each split preserves opening and connection costs, creates a complete copy, and removes at least one positive assignment from the remaining facility. Thus the process terminates in polynomial time. Discarding zero-opening and unused copies preserves feasibility and cannot increase cost, so the solution remains optimal. Reusing the notation, we obtain
$x_{ij}^*\in\{0,y_i^*\}, F_j=\{i:x_{ij}^*>0\}, \text{and} \sum_{i\in F_j}y_i^*=1.$
Every remaining copy belongs to $F_j$ for at least one client $j$.



Set $\bar y_i=\gamma y_i^*$.
Process clients one at a time. For each client $j$, order the facilities in $F_j$ by distance from $j$, breaking ties arbitrarily, and retain the closest unit of $\bar y$-mass. Split a boundary copy if necessary. 
 Denote the resulting close set by $F_j^C$ and its complement within $F_j$ by $F_j^D$. 
\begin{equation}\label{eq:rounding-masses}
\text{Then} \qquad
 \sum_{i\in F_j^C}\bar y_i=1,
 \qquad
 \sum_{i\in F_j^D}\bar y_i=\gamma-1.
\end{equation}

Whenever a boundary copy is split, both resulting copies inherit its membership in every previously defined support, close, and distant set. Thus all sets refer to a common refinement of the facility copies. Every remaining copy $i$ belongs to $F_j$ for some client $j$. If $i\in F_j^C$, the mass identities above imply $\bar y_i\le1$; otherwise $i\in F_j^D$, so
$\bar y_i\le\gamma-1<1$. Consequently, every selection
probability satisfies $0<\bar y_i\le1$.
Moreover, for every representative $s$,
$\sum_{i\in F_s^C}\bar y_i=1$, so selecting one copy
from $F_s^C$ with these probabilities is well defined.

\[ \text{Define}, \qquad 
 d_j^C=\sum_{i\in F_j^C}\bar y_i d(i,j),\qquad
 d_j^D=\frac{1}{\gamma-1}\sum_{i\in F_j^D}\bar y_i d(i,j),\qquad
 a_j=\max_{i\in F_j^C}d(i,j).
\]
Closest-first filtering and the restriction to $E_R$ imply
\begin{equation}\label{eq:rounding-distance-identities}
 d_j^C\le a_j\le d_j^D,
 \qquad a_j\le R_j,
 \qquad d_j^C+(\gamma-1)d_j^D=\gamma C_j^*.
\end{equation}
The last identity holds because $F_j^C$ and $F_j^D$ partition the scaled original facilities serving $j$.

\subsection{Clustering and a randomized rounding}

We first describe a randomized algorithm. The final algorithm fixes its
choices deterministically by conditional expectations.

\begin{enumerate}
\item Start with all clients unassigned. Choose an unassigned client $s$ of
minimum $a_s$ as a representative, assign to $s$ every remaining client $j$ with
$F_j^C\cap F_s^C\ne\varnothing$, and remove these clients from consideration.
Repeat until all clients have a representative. Let $K$ be the set of representatives and let
$\sigma(j)$ be the representative assigned to $j$.

\item Independently for each representative $s\in K$, select exactly one copy from
$F_s^C$, selecting copy $i$ with probability $\bar y_i$. Independently of these
choices and of one another, select every copy outside
$\bigcup_{s\in K}F_s^C$ with probability $\bar y_i$.

\item Open each original facility having a selected copy, merge its selected
copies, and assign each client to its closest open facility.
\end{enumerate}

\begin{lemma}\label{lem:rounding-clusters}
The sets $F_s^C$, $s\in K$, are pairwise disjoint. For every client $j$ and
its representative $s=\sigma(j)$ we have, $F_j^C\cap F_s^C\ne\varnothing$ and, $a_s\le a_j$.
\end{lemma}
\begin{proof}
When $s$ is chosen, all remaining clients whose close sets intersect $F_s^C$
are assigned and removed. No later representative can have an intersecting close
set. The client $j$ was still unassigned when $s$ was chosen, so the
minimum-$a_s$ rule gives $a_s\le a_j$.
\end{proof}

\subsection{Proof of Theorem \ref{thm:general-intro}}

To bound the cost, charge a selected copy the full opening cost of its original facility. Let $H$ denote this total copy charge plus the clients' actual connection costs. Merging can only reduce the opening charge, and therefore the cost of the returned solution is at most $H$.

\begin{lemma}\label{lem:rounding-opening-fairness}
The expected copy opening charge is $\gamma F^*$ and every rounding outcome
satisfies $d(j,S)\le3a_j\le3R_j \text{ for all } j \in C.$
\end{lemma}
\begin{proof}
The disjointness of the representative sets makes the rounding well defined, and
every copy is selected with probability $\bar y_i$. Hence the expected copy charge is
$\sum_i f_i\bar y_i=\gamma F^*$.

Fix $j$ and put $s=\sigma(j)$. Choose a copy
$\ell\in F_j^C\cap F_s^C$, and let $I_s\in F_s^C$ be the copy selected by
representative $s$. The triangle inequality gives
\[
 d(j,S)\le d(j,I_s)
 \le d(j,\ell)+d(\ell,s)+d(s,I_s)
 \le a_j+2a_s\le3a_j\le3R_j.
\]
Only $a_s\le a_j$ is needed; no ordering of $R_s$ and $R_j$ is assumed.
\end{proof}

The maximum-distance ordering and the resulting three-edge bound
are inherited from the locality analysis of Byrka, Srinivasan,
and Swamy~\cite[Section~3.4 and Corollary~3] {ByrkaSrinivasanSwamy2010}; the restricted fractional support
supplies the additional inequality $a_j\le R_j$ used here.

\begin{lemma}
\label{lem:rounding-connection}
The expected total connection cost of the randomized algorithm is at most $\gamma C^*$.
\end{lemma}

\begin{proof}
The independent rounding groups are the representative sets $F_s^C$
and the singleton copies outside their union. Each representative group
selects exactly one copy, whereas each singleton group either
selects its copy or selects nothing.

We first bound the probability that the rounding misses a set
$A$ of facility copies. For each group $B$, put
$p_B=\sum_{i\in A\cap B}\bar y_i$.
The probability that $B$ selects no member of $A$ is $1-p_B$.
Independence across groups therefore gives
\begin{equation}\label{eq:rounding-miss}
 \Pr(S\cap A=\varnothing)
 =\prod_B(1-p_B)
 \le \exp\left(-\sum_Bp_B\right)
 =\exp\left(-\sum_{i\in A}\bar y_i\right),
\end{equation}
where we identify selected copies with their original locations
when measuring distances.

Fix a client $j$ and put $L=3a_j$. Consider the following
auxiliary experiment, used only for the analysis.
Independently generate
$N_C\sim\operatorname{Poisson}(1)$ and
$N_D\sim\operatorname{Poisson}(\gamma-1)$.
Draw $N_C$ close copies, each with probability $\bar y_i$
for $i\in F_j^C$, and $N_D$ distant copies, each with
probability $\bar y_i/(\gamma-1)$ for $i\in F_j^D$.
All draws are independent of one another and of the counts.
Let $T$ be the resulting multiset and define
$
 Z_j=\min\{L,d(j,T)\}$ and, $d(j,\varnothing)=+\infty.$

A Poisson number of independent draws with mean $\lambda$
misses a set having single-draw probability $p$ with probability
$
 \sum_{k=0}^{\infty}
 e^{-\lambda}\frac{\lambda^k}{k!}(1-p)^k
 =e^{-\lambda p}.
$
Applying this identity to the close and distant draws shows that,
for every $A\subseteq F_j$, $
 \Pr(T\cap A=\varnothing)
 =\exp\left(-\sum_{i\in A}\bar y_i\right).$

For $0\le t<L$, let
$A_t=\{i\in F_j:d(i,j)\le t\}$.
If $d(j,S)>t$, then the actual rounding selects no copy in $A_t$.
Thus, by \eqref{eq:rounding-miss},
$
 \Pr(d(j,S)>t)
 \le \exp\left(-\sum_{i\in A_t}\bar y_i\right)
 =\Pr(Z_j>t).
$
For $t\ge L$, both tail probabilities are zero, since
Lemma~\ref{lem:rounding-opening-fairness} guarantees
$d(j,S)\le3a_j=L$ in every outcome.
Integrating the tails gives
$\mathbb{E}[d(j,S)]\le\mathbb{E}[Z_j]$.

To bound $\mathbb{E}[Z_j]$, charge the distance of the first
close draw if $N_C>0$, the distance of the first distant draw
if $N_C=0$ and $N_D>0$, and $L$ if both counts are zero.
This charge is at least $Z_j$ in every outcome.
The expected distance of a close draw is $d_j^C$, and that
of a distant draw is $d_j^D$. Hence
\begin{align}
 \mathbb{E}[d(j,S)]
 &\le (1-e^{-1})d_j^C
      +e^{-1}(1-e^{-(\gamma-1)})d_j^D
      +e^{-\gamma}L \notag \\
 &\le (1-e^{-1})d_j^C
      +(e^{-1}+2e^{-\gamma})d_j^D,
 \label{eq:rounding-connection}
\end{align}
where the last inequality uses $L = 3a_j$ and $a_j\le d_j^D$ from
\eqref{eq:rounding-distance-identities}.

For $\gamma=1.725$, the inequalities $ 
 e>\sum_{k=0}^{6}\frac1{k!}
   =\frac{1957}{720}>\frac{125}{46}$ and,
 $e^{69/40}>
 \sum_{k=0}^{7}\frac{(69/40)^k}{k!}>
 \frac{2000}{357}
$
give
\begin{equation}\label{eq:rounding-constant}
 e^{-1}+2e^{-\gamma}
 <\frac{46}{125}+\frac{357}{1000}
 =\frac{29}{40}
 =\gamma-1.
\end{equation}
Combining \eqref{eq:rounding-connection} with
\eqref{eq:rounding-distance-identities} therefore yields $
 \mathbb{E}[d(j,S)]
 \le d_j^C+(\gamma-1)d_j^D
 =\gamma C_j^*.
$
Summing over all clients proves the total connection-cost bound.
\end{proof}

Combining Lemma~\ref{lem:rounding-connection} with the expected
opening-charge bound in
Lemma~\ref{lem:rounding-opening-fairness}, we obtain
 $\mathbb{E}[H]
 \le\gamma F^*+\gamma C^*
 =\gamma\LP.$

%% file: exp-rewritten.tex
\section[Experiments]{Experiments\footnote{\url{https://github.com/milindmaiti/IFFL_Implementation}}}
\label{sec:experiments}

We evaluate the randomized rounding algorithm of Section~\ref{sec:rounding}
and two reference algorithms on Euclidean instances derived from real data.
We vary the client radii and measure how tightening them changes the restricted
LP bound, how the algorithms' costs compare with that bound, and how often
and by how much clients are served outside their prescribed radii.
\vspace*{-0.6cm}
\paragraph*{Instance construction}
We use two UCI data sets~\cite{adult,bank}. For Adult, we retain the six
continuous attributes of the $45{,}222$ complete records. For Bank Marketing,
we use \texttt{age}, \texttt{balance}, and \texttt{duration} from all $45{,}211$
records. We standardize each retained attribute to zero mean and unit variance
and use Euclidean distance in the resulting feature space.

For each data set and each $m\in\{200,400\}$, we sample $n=1000$ clients and $m$
candidate facilities uniformly without replacement from the records. When
sampling facilities, we exclude records whose feature vectors coincide with
those of sampled clients, so $d(j,F)>0$ for all $j\in C$.
Let $A_i$ be the set of the $25$ clients closest to facility $i$. We set
\[
  a_i=\frac{1}{25}\sum_{j\in A_i}d(i,j),
  \qquad
  \bar d=\frac{1}{n}\sum_{j\in C}d(j,F),
  \qquad
  f_i=c\,\bar d\,\frac{a_i^{-1}}{\frac{1}{m}\sum_{\ell\in F}a_\ell^{-1}},
\]
where $c\in\{4,7,15\}$. This assigns higher opening costs to facilities whose
nearby clients are closer on average. The mean opening cost is $c\,\bar d$,
so $c$ scales opening costs relative to the mean nearest-facility distance.

\paragraph*{Client radii.}
We generate radii using three schemes.
\begin{enumerate}
  \item \emph{Nearest-facility radii.}
  Set $R_j=\alpha\,d(j,F)$ for $\alpha\in\{1,1.5,2,3,4\}$.
  \item \emph{Neighborhood radii.}
  Following Jung, Kannan, and Lutz~\cite{JungKannanLutz2020}, take the
  distance from $j$ to its $\lceil n/k\rceil$-th nearest client for
  $k\in\{150,50,20,10\}$, and raise it to $d(j,F)$ if necessary.
  \item \emph{Common radii.}
  Set $R_j=\tau\max_{h\in C}d(h,F)$ for all $j$, with
  $\tau\in\{1,1.5\}$.
\end{enumerate}
Each scheme gives every client at least one admissible facility. For the two
client-specific schemes, we plot cost against the \emph{effective radius
multiplier}, defined as $\operatorname{median}_{j\in C}\bigl(R_j/d(j,F)\bigr)$.
It equals $\alpha$ for the nearest-facility scheme. In the pooled cost plot,
the four neighborhood settings have multipliers of approximately $1.36$,
$1.92$, $2.67$, and $3.40$.
The four base instances, three opening-cost scales, and eleven radius
settings yield $132$ configurations. The comparisons below focus on the
$108$ configurations with client-specific radii.

\begin{figure}[tb]
  \centering
  \includegraphics[width=\textwidth]{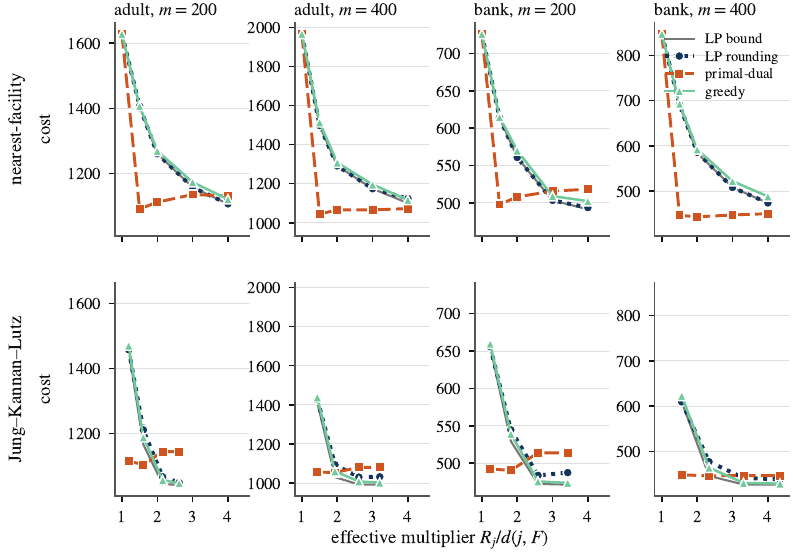}
  \caption{Total solution costs and restricted LP values at $c=7$, plotted
  against the effective radius multiplier. Each column shows one instance, with a common horizontal axis throughout.
  The two radius schemes share a vertical scale within each column.}
  \label{fig:absolute}
\end{figure}
\vspace*{-0.5cm}
\paragraph*{Algorithms and evaluation}
Let $\mathrm{LP}$ be the optimum of the restricted relaxation from
Section~\ref{sec:rounding} for the current configuration. For an open
facility set $S$, the normalized cost is $\operatorname{cost}(S)/\mathrm{LP}$.
Since $\mathrm{LP}\le \mathrm{OPT}_{\mathrm{FL}}$, this ratio upper-bounds
$\operatorname{cost}(S)/\mathrm{OPT}_{\mathrm{FL}}$, with equality whenever
the LP has an integral optimum. To measure radius violations, we record
$d(j,\sigma(j))/R_j$ for each client $j$ served by facility $\sigma(j)$ and
report the empirical complementary cumulative distribution of these ratios.

\textsc{LP-Rounding} applies the randomized algorithm of
Section~\ref{sec:rounding}. We average its cost over $50$ independent
roundings per configuration. The analysis bounds its expected cost
by $1.725\,\mathrm{LP}$ and guarantees a service distance of at most
$3R_j$ for each client $j$ in every outcome.

\textsc{Greedy} restricts the greedy facility-location algorithm~\cite[Algorithm
9.2]{williamson-shmoys} to admissible edges. At each step it chooses a facility
and a set of unserved clients that minimize the net cost per newly served
client. The net cost includes any new opening cost and the new connection
costs, less the savings from reassigning previously served clients. Both
assignments and reassignments use admissible edges, so the solution is
strictly fair.

\textsc{Primal-Dual} uses Jain--Vazirani dual growth~\cite{JainVazirani2001}
with client contributions restricted to admissible edges. The dual variables
of unconnected clients grow at a common rate, and a facility opens
provisionally once its opening cost is covered. We retain a maximal subset
of the provisional facilities with pairwise disjoint sets of contributing
clients. For common radii, this adaptation gives a
$(3,3)$-approximation~\cite{ArutyunovaEtAl2024}.

\begin{figure}[tb]
  \centering
  \begin{subfigure}[b]{0.48\textwidth}
    \centering
    \includegraphics[width=\linewidth]{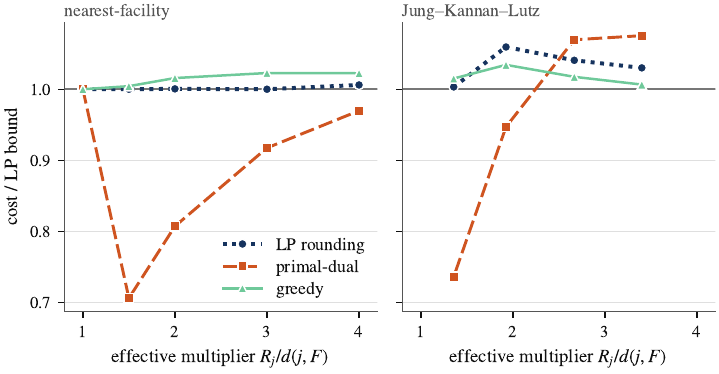}
    \caption{Cost relative to the restricted LP bound.}\label{fig:cost}
  \end{subfigure}\hfill
  \begin{subfigure}[b]{0.48\textwidth}
    \centering
    \includegraphics[width=\linewidth]{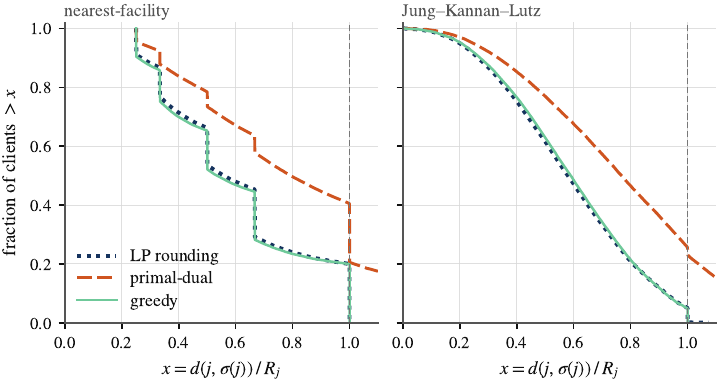}
    \caption{Radius violations across clients.}\label{fig:fairness}
  \end{subfigure}
  \caption{Comparison under the two client-specific radius schemes.
  (\subref{fig:cost}) Mean cost ratio for each radius setting, averaged
  over the four instances and three opening-cost scales and plotted
  against the effective radius multiplier.
  (\subref{fig:fairness}) Empirical fraction of clients with
  $d(j,\sigma(j))/R_j>x$, pooled across configurations within each
  scheme and truncated at $x=1.1$.}
  \label{fig:cost-fairness}
\end{figure}

\vspace*{-0.5cm}
\paragraph*{Effect of tightening the radii.}
Tightening the radii restricts the admissible edge set and can only increase the
restricted LP value. At $c=7$, the LP value at $\alpha=1$ is between
$1.47$ and $1.79$ times its value at $\alpha=4$ across the four instances
(Figure~\ref{fig:absolute}). For the neighborhood scheme, the ratio between
the tightest and loosest settings ranges from $1.39$ to $1.43$. The tightest
neighborhood setting yields a lower LP value than $\alpha=1$ on each
instance. We observe larger changes in the LP value at higher opening-cost
scales.
\vspace*{-0.5cm}
\paragraph*{Cost relative to the LP bound.}
\textsc{Greedy} has a mean cost ratio of $1.015$ and a maximum of $1.082$
(Figure~\ref{fig:cost}). Thus its strictly fair solutions cost at most
about $1.5\%$ above the strictly fair optimum on average across these
configurations. In contrast, Theorem~\ref{thm:hardness-intro} rules out
a sublogarithmic worst-case cost guarantee for deterministic polynomial-time
algorithms that satisfy every radius, unless $\mathrm{P}=\mathrm{NP}$; that bound is realized by instances encoding Set Cover rather than by geometric data.
For the $60$ configurations in which the LP solver returns an integral
optimum, \textsc{LP-Rounding} attains the LP bound and satisfies every
radius. Its mean cost ratio on the remaining $48$ configurations is
$1.035$. The minimum observed cost ratio for \textsc{Primal-Dual} is $0.474$.
A ratio below one implies a radius violation because every strictly fair
solution costs at least $\mathrm{LP}$.
\vspace*{-0.5cm}
\paragraph*{Radius violations.}
\textsc{Greedy} satisfies every radius by construction. \textsc{LP-Rounding}
serves at least $99.8\%$ of client assignments within the prescribed radius
under either scheme, even though its guarantee only bounds the service
distance by $3R_j$, and its largest observed ratio is $1.474$.
\textsc{Primal-Dual} exceeds the prescribed radius for $20.5\%$ of client
assignments under nearest-facility radii and $23.1\%$ under neighborhood
radii, the heights of its curve at $x=1$ in Figure~\ref{fig:fairness}. The upper tails for \textsc{Primal-Dual}
fall outside the plotted range and differ markedly between the schemes.
Under neighborhood radii, the largest ratio is $2.22$. Under nearest-facility
radii, $2.7\%$ of client assignments exceed $3R_j$, and the largest ratio is
$216$.

%% file: Conclusion.tex
\section{Conclusion}

Our results identify the radius relaxation needed for constant-cost approximation of individually fair facility location by deterministic algorithms running in polynomial time. A violation strictly below three retains Set Cover hardness, even with equal opening costs and uniform radii. At factor three, the LP-rounding algorithm achieves cost at most $1.725$ times the optimum satisfying the original radii, for arbitrary
client radii. The optimal cost approximation factor at this radius threshold remains open. Extending this combination of cost and radius guarantees to outliers, penalties, or capacity constraints requires additional rounding and reassignment arguments and are interesting open research directions. Since papers build on existing results, it would be critical to quickly correct the implications of the claimed results by Liu et al \cite{LiuXuZhang2026}.


\section{Statement on the Use of AI Tools}
ChatGPT (GPT-6.1 Sol Ultra) was used to assist with partially drafting the manuscript and technically verifying the results and mathematical arguments. Claude was used to generate code for the implementation of the proposed algorithm. The authors take full responsibility for the content and correctness of the submitted work.

%% file: derandomization.tex
\section{Derandomization}
\label{sec:deram}

We derandomize the finite group choices directly. This also avoids any
need to sample Poisson random variables, which were introduced only to
prove Lemma~\ref{lem:rounding-connection}.

\begin{lemma}\label{lem:rounding-derandomization}
One can choose the rounding outcome deterministically in polynomial time
so that $H\le\mathbb{E}[H]$.
\end{lemma}
\begin{proof}
Fix any positive-probability outcomes for some of the groups. The remaining
groups are still independent and retain their original distributions.
For client $j$, let $\lambda_j$ be its distance to the closest copy selected
by the fixed groups, or $+\infty$ if none has been selected. For an unfixed
group $B$, put
\[
 q_{Bj}(t)=1-\sum_{\substack{i\in B\\d(i,j)\le t}}\bar y_i.
\]
Writing $D_j$ for the eventual connection distance, independence gives the
exact conditional probability
\begin{equation}\label{eq:rounding-conditional-tail}
 \Pr(D_j>t\mid\text{fixed outcomes})
 =\mathbf{1}_{\{\lambda_j>t\}}
   \prod_{B\text{ unfixed}}q_{Bj}(t).
\end{equation}
The formula remains valid after copies are merged, since copies of the
same original facility have identical distances to every client.

Let $\delta_1<\cdots<\delta_h$ be the distinct distances from $j$ to all
copies. At least one representative exists, and each representative group always selects a
copy, so $D_j$ is finite in every possible completion. Its conditional
expectation is therefore exactly
\begin{equation}\label{eq:rounding-conditional-mean}
 \delta_1+\sum_{r=1}^{h-1}(\delta_{r+1}-\delta_r)
   \Pr(D_j>\delta_r\mid\text{fixed outcomes}).
\end{equation}
The conditional expected copy opening charge is the charge already fixed,
plus $\sum_{B\text{ unfixed}}\sum_{i\in B}f_i\bar y_i$.
Equations~\eqref{eq:rounding-conditional-tail}--\eqref{eq:rounding-conditional-mean}
thus compute the conditional expectation of $H$ using rational arithmetic.

Process the groups one at a time. For the next group, evaluate this
conditional expectation under each of its positive-probability outcomes
and choose an outcome attaining the minimum. The current expectation is
the probability-weighted average of these values, so it cannot increase.
After all groups have been fixed, $H$ is at most its original expectation.

There are polynomially many copies, groups, outcomes, and distance
thresholds. The LP has a rational optimal solution of polynomial encoding
length, and splitting and scaling preserve polynomial encoding length.
The formulas above use sums of polynomially many products, each having
polynomially many rational factors. Their exact numerators and denominators
therefore have polynomial encoding length. No division by the probability
of a conditioning event is needed: fixed groups are simply removed from
the product. The entire procedure runs in polynomial time.
\end{proof}


%% file: badExampleforDualFitting.tex
\section{An explicit failure of the claimed dual scaling}\label{sec:correction}
The proof of Theorem~1 in~\cite{LiuXuZhang2026}, page~9, scales the final client values $\widehat\alpha_j$ by $1/2$ and asserts the facility constraints
\begin{equation}\label{eq:false-scaling}
 \sum_{j:(i,j)\in E_{\rm fair}}
 \left(\frac{\widehat\alpha_j}{2}-d(i,j)\right)_+\leq f_i,
 \qquad E_{\rm fair}=\{(i,j):d(i,j)\leq R_j\}.
\end{equation}
Here $(z)_+=\max\{z,0\}$. The justification invokes the metric facility-location dual-fitting analysis and asserts that deleting inadmissible edges preserves it. The following instance violates both~\eqref{eq:false-scaling} and the claimed cost factor. We analyze the contributions and reassignment rules stated in that paper directly.

\begin{figure}[t]
\centering
\begin{tikzpicture}[x=1cm,y=1cm,font=\small]
\draw[rounded corners] (0,0) rectangle (2.7,3.7);
\draw[rounded corners] (3.2,0) rectangle (5.9,3.7);
\node at (1.35,4.0) {Column $A$};
\node at (4.55,4.0) {Column $B$};
\foreach \yy/\sz/\ll in {3.3/16/1,2.6/8/2,1.9/4/3,1.2/2/4,0.5/1/5}{
 \fill[blue!8] (0.12,\yy-0.23) rectangle (5.78,\yy+0.23);
 \node at (1.35,\yy) {$\sz$ \ifnum\sz=1 element\else elements\fi};
 \node at (4.55,\yy) {$\sz$ \ifnum\sz=1 element\else elements\fi};
 \node[anchor=west] at (6.15,\yy) {$L_{\ll}$};
}
\end{tikzpicture}
\caption{The Set Cover instance underlying the finite counterexample. Each element becomes two clients in the metric. The published algorithm opens the five layers; the two columns are optimal. The shading indicates set membership, not geometric distance.}\label{fig:layers}
\end{figure}
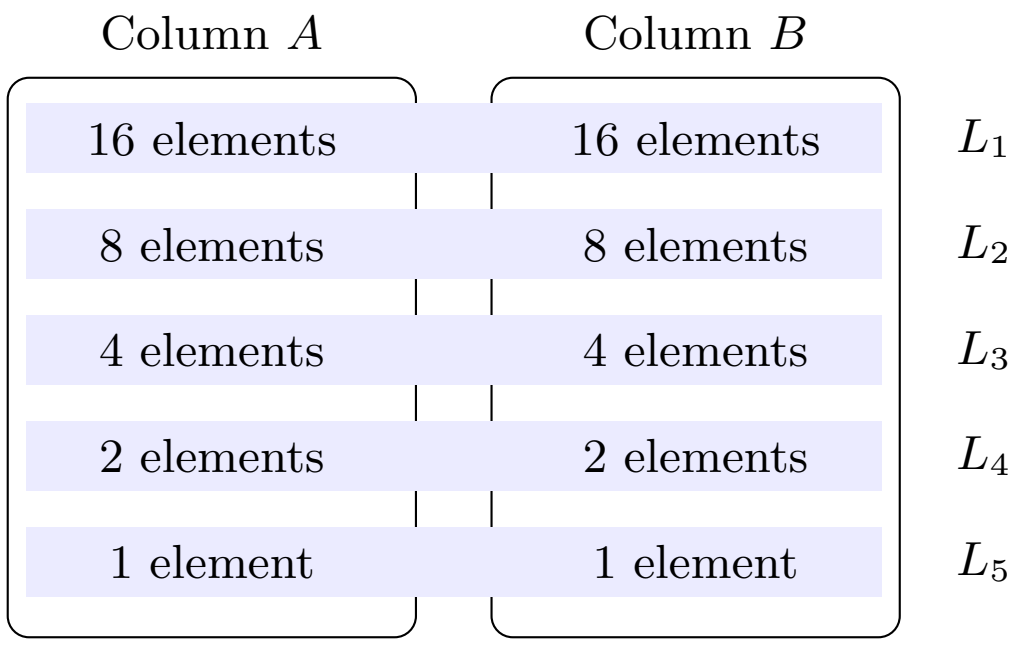

Take two disjoint sets of elements $A$ and $B$, each of size $31$. Partition each into five groups,
\[
 A=\mathop{\dot\bigcup}_{\ell=1}^5 G^A_\ell,
 \qquad B=\mathop{\dot\bigcup}_{\ell=1}^5 G^B_\ell,
 \qquad |G^A_\ell|=|G^B_\ell|=2^{5-\ell}.
\]
The seven available sets are the two columns $A,B$ and the five layers $L_\ell=G^A_\ell\cup G^B_\ell$. Apply the metric construction above, with $q=62$, $N=124$, and $M=62^2=3844$. Figure~\ref{fig:layers} illustrates the set family.

\begin{proposition}\label{prop:counterexample}
On this feasible uniform-radius instance, the facility-location algorithm of~\cite{LiuXuZhang2026} opens all five layer facilities and has cost ratio $52/21>2$. For each column facility, the left side of~\eqref{eq:false-scaling} equals $4774$, exceeding its opening cost $3844$.
\end{proposition}
\begin{proof}
Every admissible connection has length one. Once a client has been served, its connection distance remains one, so it offers no positive reassignment savings to any other admissible facility. All active clients have the common growing value $\widehat\alpha_j=t$. An unopened facility incident to $a$ active clients therefore receives contribution $a(t-1)$ for $t\geq1$.

Immediately before layer $L_\ell$ opens, its number of active clients is
\[
 q_\ell=2^{7-\ell}\in\{64,32,16,8,4\}.
\]
Each column has $q_\ell-2$ active clients, and each later layer has at most $q_\ell/2$. The unique next tight facility is consequently $L_\ell$, at time
\[
 t_\ell=1+\frac{M}{q_\ell}.
\]
At that time a column's contribution is $M(q_\ell-2)/q_\ell<M$, so neither column opens. The layers open in order and all clients are served at distance one. No available set covers the entire universe, while the two columns cover it. Thus
\[
 \frac{\mathrm{ALG}}{\OPT_{\rm FL}}
 =\frac{5M+124}{2M+124}=\frac{52}{21}>2.
\]
Within either column, precisely $q_\ell/2$ clients freeze at value $t_\ell$ for each layer. All terms in the proposed truncated sum are positive, and their sum is
\[
 \sum_{\ell=1}^5\frac{q_\ell}{2}
 \left(\frac{1+M/q_\ell}{2}-1\right)
 =\frac{5M}{4}-\frac14\sum_{\ell=1}^5q_\ell
 =\frac{5M}{4}-31=4774>M.
\]
This proves the violation independently of tie breaking and independently of any complexity assumption.
\end{proof}

The example isolates the failed assertion: preservation of the feasible set under admissible-edge restriction is valid, but preservation of the metric dual-fitting facility inequality is not. Increasing the number of layers gives the same behavior with an unbounded logarithmic ratio.